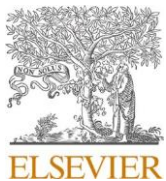

Contents lists available at ScienceDirect

# Sensors and Actuators: A. Physical

journal homepage: www.journals.elsevier.com/sensors-and-actuators-a-physical

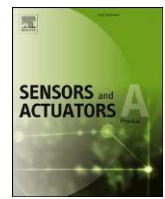

# Single-step laser patterning and thinning of biocompatible MEMS flow sensor

Mohammad Nizar Mohamed Zukri , Muhammad Salman Al Farisi , Yoshihiro Hasegawa , Mitsuhiro Shikida

*Department of Biomedical Information Sciences, Hiroshima City University, Hiroshima 731-3194, Japan*

### ABSTRACT

Micro-electro mechanical systems (MEMS) thermal flow sensors are increasingly used for compact, low-power flow monitoring in biomedical applications. However, silicon-based method for sensor fabrication is limited by high cost, rigidity, and multi-step cleanroom processes. This study presents a single-step fiber laser micromachining method for fabricating biocompatible, free-standing MEMS thermal flow sensors from ultrathin titanium foil. The process combines patterning and localized thinning in single-step process, with titanium serving as resistive sensing element. A dual-matrix optimization approach consisting of a Threshold Mapping Matrix (TMM) and Energy Density Matrix (EDM) was used to determine optimized parameters without repeated trial-and-error. For localized thinning, sequential R-T scans with cooling intervals reduced redeposition from the Gaussian beam profile and produced uniform thickness reduction from 50 μm to 20–30 μm. The fabricated sensors were evaluated through thermal coefficient resistance (TCR) measurement, Infrared (IR) thermography, and airflow tests under steady and cyclic conditions controlled by artificial ventilation system. The fabricated devices showed a stable TCR of 3278 ppm °C⁻¹, a linear relationship calibration curve between velocity and resistance with $R^2 = 0.986$ and a 54% improvement in thermal response was achieved with the free-standing structure design compared to substrate-fixed designs. This fabrication approach removes the need for photolithography, wet/dry etching, and wafer bonding, enabling faster and lower-cost production of flexible, biocompatible flow sensors. The method can be applied to other MEMS devices that require compact size, flexibility, localized thinning and free-standing structures.

## 1. Introduction

The global MEMS sensor market stood at approximately USD 18.76 billion in 2025 and is projected to grow at a compound annual growth rate (CAGR) of around 9.17% through 2030 [1]. In parallel, MEMS flow sensor segments including ultrasonic and Coriolis types are projected to reach over USD 12 billion by 2030, with a CAGR of about 5.7% [2]. These projections reflect the growing demand for Industry 4.0, automotive, agriculture, and healthcare sectors, where miniaturization trends are driving product development in compact factors, more accurate and reliable sensing solutions [3][4][5].

Within this context, MEMS thermal-based flow sensors attain significant due to their compact size, low power usage, and rapid response capabilities [6][7]. Their sensitivity to convective cooling enables accurate real-time flow measurements, making them attractive for integration into portable and low-latency systems particularly in respiratory monitoring applications [8][9][10].

Despite their advantages, conventional MEMS fabrication methods such as photolithography, dry/wet etching, and wafer bonding are constrained by cleanroom dependency, use of rigid substrates, and high cost and complexity, limiting adaptability to flexible or wearable platforms [11][12]. These limitations have triggered interest in maskless microfabrication techniques that reduce complexity and widen substrate options. Laser micromachining has appeared as a strong candidate, offering direct patterning capability, feature sizes less than 100 μm, and straightforward patterning without lithographic masks [13][14]. Yb-fiber lasers are widely used for precision cutting and local material removal in thin metallic foils due to their beam quality, electrical-to-optical efficiency, and ease of integration into industrial systems [15][16].

Department of Biomedical Information Sciences, Hiroshima City University, Hiroshima 731-3194, Japan.
Corresponding author.
*E-mail addresses:* dj65001@e.hiroshima-cu.ac.jp (M.N.M. Zukri), alfarisi@hiroshima-cu.ac.jp (M.S. Al Farisi).

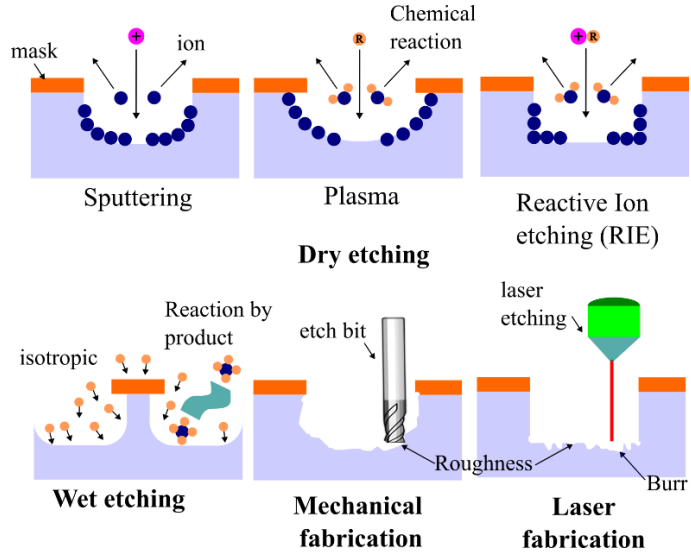

**Fig. 1.** Common fabrication techniques for material thinning and micro structuring.

Optimizing laser parameters for thin foil micromachining remains challenging. Trial-and-error tuning of power, speed, hatch spacing, and pulse repetition is still common but is inherently time-consuming and material dependent [17][18]. Process mapping approaches have improved reproducibility, but it may not capture all parameter interactions or environmental effects. Recently, machine learning (ML) and artificial intelligence (AI) techniques have been explored to accelerate laser process optimization [19][20][21], but challenges such as small datasets, limited transferability across materials, and insufficient physics-informed constraints restrict their practical adoption, especially for ultrathin metallic structures [22].

Beside the fabrication of overall structure, another aspect of MEMS-based sensing devices is the thinned diaphragm structures. These structures are extensively used as a sensing element in various MEMS devices, such as pressure sensors, microphones, and flow sensors [23][24]. For instance, in MEMS pressure sensors, a thinned diaphragm undergoes deflection due to pressure changes, altering resistance and piezoelectric signals [25][26]. In MEMS flow sensors, the diaphragm structure works to thermally isolate the free-standing sensing structure, which is Joule-heated, and airflow modulates its resistance by modifying the thermal profile [27]. Additionally, MEMS microphones rely on thinned diaphragms to detect sound-induced vibrations, converting mechanical energy into electrical signals [28]. The widespread application of thin diaphragms highlights its significance in MEMS-based sensing and actuation technologies [29].

Several methods have been established for fabricating locally thinned diaphragm structures as illustrated in Fig.1. Dry etching techniques, such as reactive ion etching (RIE) and plasma etching [30], provide high precision but require sophisticated equipment and long processing times [31]. Wet etching offers isotropic material removal with good uniformity but lacks control over etch depth [32]. Mechanical thinning using tool bits presents limitations at the microscale [33]. Laser micromachining has emerged as a promising alternative, with various laser sources such as $CO_2$ (10.6 μm), Nd:YAG/fiber (~1.06 μm), and the 355 nm third harmonic of Nd:YAG, while excimer lasers (e.g., 248–351 nm) provide efficient ablation of polymers and thin films [34][35]. For metal structures, fiber and YAG lasers are suitable due to their wavelength compatibility [36]. In Processing metallic foils at micrometer scale presents known difficulties, including spatter and resolidified debris (redeposition), heat-affected zones (HAZ), and non-uniform ablation depth caused by the Gaussian beam profile and scan-path overlap as illustrated in Fig. 2 [37][38]. These effects depend strongly on scanning strategy (hatch spacing, overlap ratio, loop count), average power, pulse energy, and repetition rate, motivating the development of explicit process maps to identify stable regimes for cutting, thinning, and surface finishing [39]. Moreover, using multiple passes can improve both depth control and surface finish [40][41].

For biomedical and implantable applications, material choice is constrained by biocompatibility and corrosion resistance. Titanium remains as a benchmark for long-term tissue contact owing to its stable oxide layer, corrosion resistance, and low ion release, and it is already widely adopted in implantable medical devices [42][43]. From a sensing standpoint, thermal flow elements typically employ resistive materials with a positive temperature coefficient of resistance (TCR) and adequate sheet resistance: platinum is the canonical resistance temperature detector (RTD) material (α ≈ 0.00385 K⁻¹, IEC 60751), while nichrome alloys are valued for stability and tunable TCR, and are frequently used in microheaters and thermistors [44][45]. Recent work has demonstrated that Ni-based thin films and titanium foils can be directly patterned by laser micromachining to produce flexible RTD structures, confirming the compatibility of laser-based digital patterning with biocompatible sensing elements [46][47].

In this work, we address these gaps by developing a single-step fiber laser micromachining process that integrates patterning and selective thinning of biocompatible metallic foils to fabricate free-standing MEMS thermal flow sensors. The approach combines an interdigitated design to achieve the target resistance within a compact footprint, a dual-matrix optimization framework for robust penetration and scan parameter selection, and a sequential thinning strategy to mitigate redeposition effects while preserving mechanical integrity. Device performance is validated through thermal imaging and airflow measurements under steady and cyclic flow conditions, demonstrating the method's suitability for biomedical respiratory monitoring and other applications requiring flexible, high-sensitivity MEMS flow sensors.

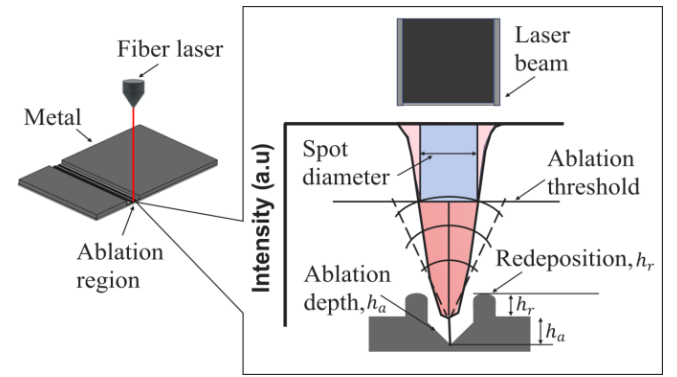

**Fig. 2.** Gaussian Beam Profile, characterized by its bell-shaped intensity distribution.

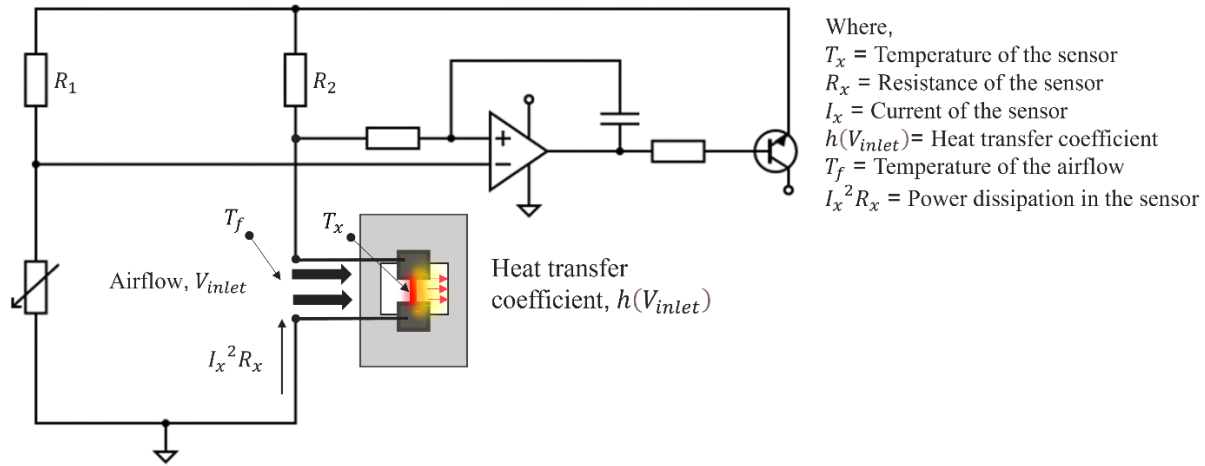

Where,
$T_x$ = Temperature of the sensor
$R_x$ = Resistance of the sensor
$I_x$ = Current of the sensor
$h(V_{inlet})$= Heat transfer coefficient
$T_f$ = Temperature of the airflow
$I_x{}^2 R_x$ = Power dissipation in the sensor

**Fig. 3.** Schematic of an in-line thermal anemometer for flow sensors detection.

## 2. Principle of Operation

### 2.1 Thermoresistive Flow Sensing Mechanism

The sensing principle is based on thermal anemometry in constant temperature anemometer (CTA) mode as illustrated in Fig. 3 [48][49], where the sensing element is a thermoresistive material heated by Joule heating. When an electric current $I_x$ flows through the resistive element, electrical power is dissipated as Joule heating according to Eq (1):

$$P = I_x{}^2 R_x \tag{1}$$

Here, $R_x$ is the electrical resistance of the sensing element. Airflow over the sensor enhances convective heat transfer, increasing the heat dissipation rate and thus affecting the sensor's electrical equilibrium. In CTA mode, the sensor's resistance $R_x$ and temperature $T_x$ are maintained constant, while the current $I_x$ is adjusted to balance electrical heating with convective cooling. The equilibrium is expressed as Eq (2):

$$I_x{}^2 R_x = h(V_{inlet}) . A_s (T_x - T_f) \tag{2}$$

Where $h(V_{inlet})$ is the heat transfer coefficient (W/m²K) dependent on airflow velocity $V_{inlet}$, $A_s$ is the effective heat transfer surface area, $T_x$ is the sensor temperature, and $T_f$ is the airflow temperature.

The electrical resistance of the sensing element varies with temperature according to the Temperature Coefficient of Resistance (TCR) as Eq (3):

$$R(T) = R_0[1 + \alpha(T - T_0)] \tag{3}$$

Where $R_0$ is the baseline resistance at temperature $T_0$, and $\alpha$ is the TCR. This relationship enables indirect measurement of airflow through current variation in CTA mode, as increased velocity raises $h$, requiring higher $I_x$ to maintain constant $T_x$ and $R_x$ [50][51][52].

Fig. 4 illustrates this mechanism: in the "no airflow" condition, the heat generated by the sensing element (heater) forms a symmetric thermal profile around the structure, maintaining a stable temperature $T_x = T_0$ and resistance $R(T) = R_0$. In the "airflow" condition, convective cooling increases, causing a temperature drop ($T_x < T_0$), which reduces the resistance $R(T)$ according to Eq. (3). In a voltage-driven configuration ($V = I_x R(T)$), this reduction in resistance leads to a lower output voltage ($V_1 < V_2$). In CTA mode, the system compensates for this by increasing the heating current $I_x$ to restore $T_x$ and $R_x$ to their set values. This process creates a direct correlation between the airflow velocity and the additional current required to maintain constant temperature, as visualized by the downstream displacement and elongation of the heated zone in the figure.

### 2.2 Interdigitated Design and Free-Standing Architecture

To achieve sufficient working resistance in a compact sensing area, the metallic foil is patterned into an interdigitated geometry, increasing the electrical path length without enlarging the overall footprint. Multiple parallel lines are connected in series, allowing resistance adjustment via line length and width while maintaining structural stability.

In this study, three sensing structures were patterned. In previous work using nichrome [53], an I-shaped design achieved functional performance because nichrome's high material resistivity, valued at 110 x$10^{-8}$ $\Omega$·m, produced approximately 6 $\Omega$ resistance within the same footprint. However, nichrome is not biocompatible, so for biomedical applications, we challenged to use titanium, which is biocompatible but has a lower resistivity (42 x$10^{-8}$ $\Omega$·m) [54]. With the same I-shaped geometry in titanium, the resistance was insufficient for thermal sensing.

Here we adopted an interdigitated structure aimed at achieving up to 10 $\Omega$ resistance while maintaining structural rigidity. The design consists of five parallel line arrays, each mechanically anchored to a support frame. This configuration increases both effective path length and surface area, improving electrical resistance and thermal exchange, as described by Pouillet's law as presented in Eq (4):

$$R = \rho\left(\frac{L}{A}\right) \tag{4}$$

where $\rho$ is the material resistivity, $L$ is the effective length of the conductor, and A is its cross-sectional area. The interdigitated design thus allows for improved thermal responsiveness without enlarging the device dimensions. Furthermore, localized laser thinning was applied in the sensing region to reduce the cross-sectional area, $A$, decreasing the thermal mass and enhancing sensitivity to transient airflow-induced temperature changes. The sensing region is suspended by removing the underlying silicone rubber, forming a free-standing structure that is thermally isolated from the substrate.

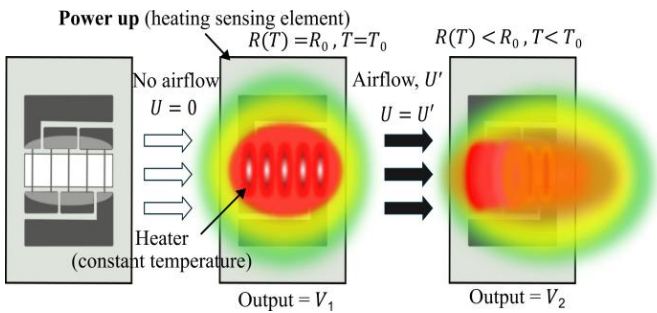

**Power up** (heating sensing element)

$R(T) = R_0, T = T_0$      $R(T) < R_0, T < T_0$

No airflow
$U = 0$

Airflow, $U'$
$U = U'$

Heater
(constant temperature)

Output = $V_1$      Output = $V_2$

**Fig. 4.** A schematic drawing of the thermal anemometer measurement principle.

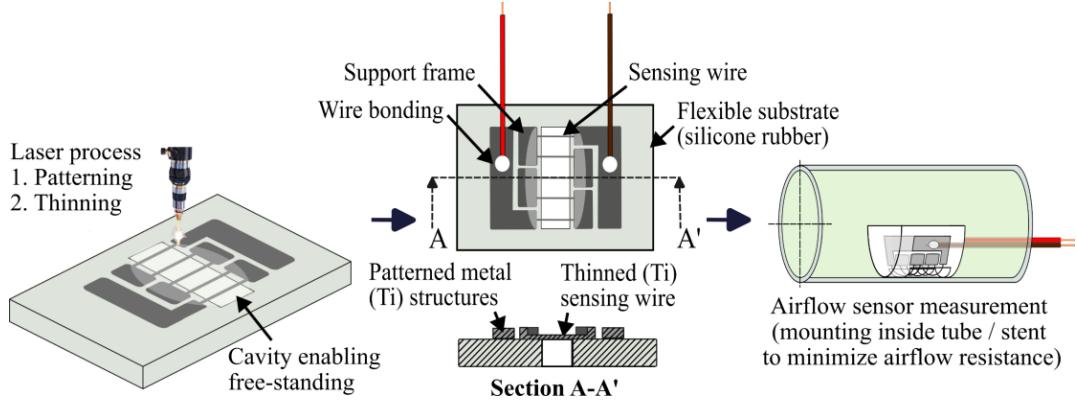

**Fig. 5.** Proposed and conceptual of 3D-shaped free-standing structures on flexible substrate for MEMS thermal flow sensor.

## 3. Materials and Methods

### 3.1 System integration and Fabrication Process

The sensing element was conceived as a free-standing structure patterned from a 50 μm-thick titanium (Ti) foil. The foil (purity > 99.6%) was commercially sourced from Nilaco Corporation (Tokyo, Japan). This thickness was selected as a practical substrate choice that provides an effective balance between flexibility and mechanical robustness, consistent with our previous work on biocompatible Ti-based stent structures [55]. In its original concept, the geometry itself is able to fold and hold shape much like an origami sheet, allowing the wire and frame to remain suspended without relying on an underlying substrate. The origami-inspired design principle opens opportunities for future integration into biomedical applications such as stents or temperature probes, where folding and deployment within confined geometries can be advantageous. The overall concept, illustrated in Fig. 5, was realized through two main laser-processing steps: (1) precision patterning to define the sensing geometry, and (2) localized thinning at the sensing region to enhance the electrical resistance and thermal sensitivity of the Ti wire.

For demonstration, a silicone rubber sheet was used as a handling substrate. Its primary role was to provide mechanical support during integration and measurement. To highlight the suspended nature of the Ti sensing element, a central cavity was laser-cut into the supporting layer, leaving the patterned wire freely suspended. This configuration not only reflects the advantage of the origami-inspired Ti design but also minimizes the thermal mass of the device. With reduced surrounding material, the sensing element was able to heat and cool more rapidly, thereby enhancing sensitivity and improving response time during airflow measurements.

To evaluate airflow sensing performance, the fabricated device was installed inside a cylindrical test tube with an inner diameter of 6.5 mm. For testing, the sensor was temporarily mounted along the tube wall, positioning the free-standing wire across the airflow path. Owing to the thin titanium architecture and origami inspired geometry, the sensing structure could be gently folded to conform to the tube curvature without mechanical damage. The measured resistance changed only slightly when bent, originating primarily from the small increase in electrical path length rather than from strain induced deformation. This confirms that the structure maintains electrical stability when integrated into curved or confined environments.

The fabrication flow is depicted in Fig. 6. The silicone sheet was pre-cut a central cavity to expose the suspended region of the sensing element. A thin layer was then prepared using a silylated urethane resin based, solvent-free adhesive (Bond Ultra-Versatile S.U. Premium Soft Adhesive, Konishi Co., Japan). To form an ultrathin and uniform coating, a small piece of silicone rubber was first used as an applicator. The adhesive was placed on the small silicone pad and then lightly tapped several times onto the main silicone substrate to transfer and spread the adhesive evenly into a very thin layer. The titanium foil was subsequently placed on top of this coated substrate. Laser patterning was performed to define the sensing geometry and supporting frame across the central cavity region. During this step, the laser separated the unwanted titanium areas from the designed sensing frame. After processing, the unwanted foil sections were peeled off manually, while the patterned sensing structure remained firmly attached to the silicone substrate through the thin adhesive layer. This configuration provides sufficient mechanical support and flexibility for subsequent installation into the test tube. The silicone base therefore functions only as a temporary demonstrator support.

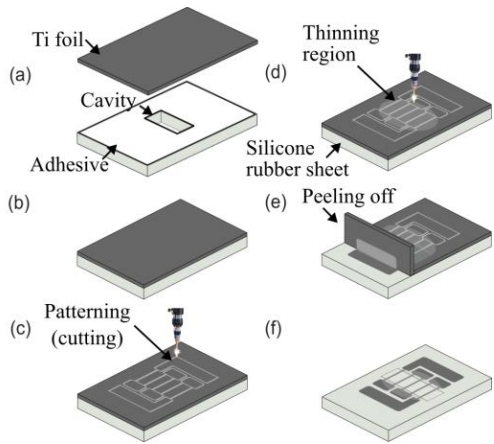

**Fig. 6.** Fabrication process: (a) applied adhesive; (b) attached Ti foil; (c) patterning the structure; (d) thinning down the sensing structure; (e) peeling off excess removal; (f) final structure.

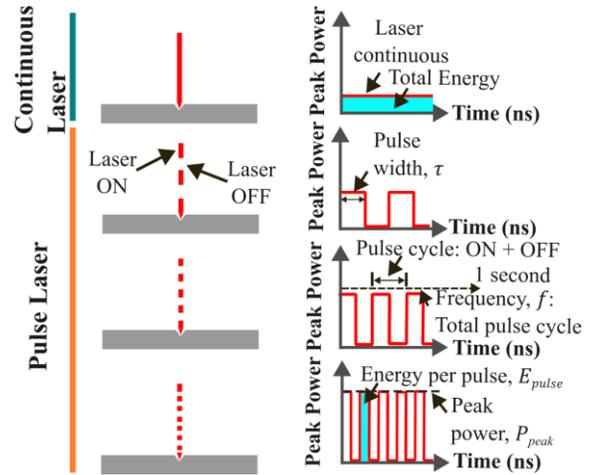

**Fig. 7.** Illustration of continuous and pulsed laser principles, highlighting laser modulation parameters.

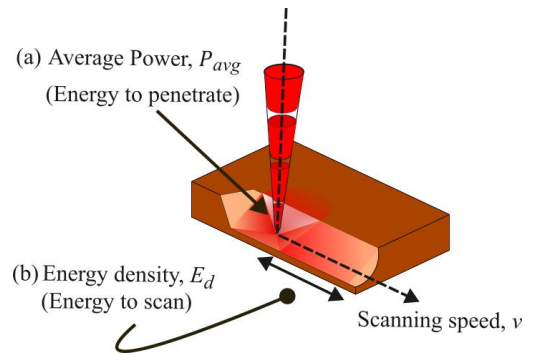

**Fig. 8.** Conceptual diagram and mathematical formula showing laser energy for (a) penetration and (b) scanning.

### 3.2 Laser Micromachining Setup

The micromachining was performed using a fiber laser system (LM110M, SmartDIYs Co., Ltd.) with a nominal spot size of 50 μm. The laser was operated in pulsed mode, which allowed independent control of peak power, pulse width, repetition frequency, and scanning speed. This flexibility enabled adjustment of both penetration depth and surface quality. The Gaussian intensity distribution of the focused beam produces higher energy density at the center of the spot, which can result in redeposition and non-uniform thinning if not properly controlled. Careful selection of pulse parameters and scanning strategy was therefore necessary to balance cutting precision with surface integrity.

The principle of pulsed laser operation is summarized in Fig. 7. Unlike a continuous laser that delivers constant power, a pulsed laser provides discrete energy packets within defined pulse widths. The total energy delivered is determined by the peak power, pulse width, and repetition frequency. Shorter pulses with high peak power favor material ablation, while longer pulse durations and higher duty cycles increase the risk of excessive heating. Fig. 8 highlights the relationship between average power and energy density during scanning. In pulsed operation, the average laser power ($P_{avg}$) is determined by the product of peak power ($P_{peak}$), pulse width ($\tau$), and repetition frequency ($f$):

$$P_{avg} = P_{peak} \times \tau \times f \qquad (5)$$

Where $P_{peak}$ is expressed as a fraction of the maximum laser output (20 W). The energy density ($E_d$) of each condition is defined as:

$$E_d = \frac{P_{avg}}{v} \qquad (6)$$

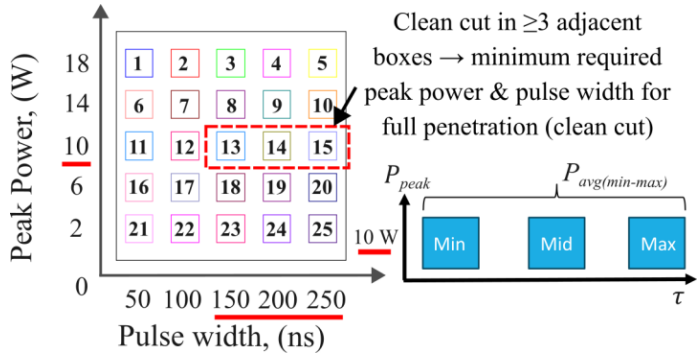

**Fig. 9.** Schematic of the Threshold Mapping Matrix (TMM) methodology.

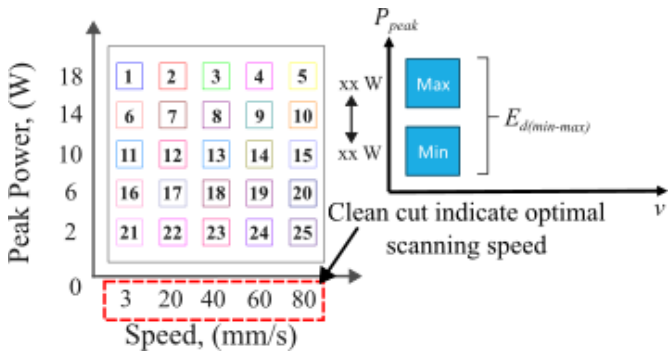

**Fig. 10.** Schematic of the Energy Density Matrix (EDM) methodology.

With $v$ representing the scanning speed. While $P_{avg}$ primarily governs the penetration capability of the beam, $E_d$ determines the effective energy delivered per unit length along the scan. By maintaining constant average power and tuning the scanning speed, an optimal energy window was identified that minimized excessive heat accumulation and reduced surface roughness. These coupled relationships formed the basis for defining stable cutting thresholds and refining thinning conditions in ultrathin titanium foils.

### 3.3 Patterning Optimization via Dual-Matrix Framework

#### 3.3.1 Threshold Mapping Matrix (TMM) and Energy Density Matrix (EDM)

To establish reliable cutting conditions for ultrathin metal foils, preliminary trials were conducted on several candidate materials including titanium, copper, and nichrome. Because the thermal conductivity and reflectivity of each foil significantly influence the cutting threshold, a systematic mapping approach was adopted. A Threshold Mapping Matrix (TMM) was constructed as a $5 \times 5$ grid, varying peak power along the y-axis and pulse width along the x-axis (Fig. 9).

Baseline trials employed a pulse frequency of 5 kHz, identified from preliminary experiments as the optimum baseline for achieving high pulse energy and stable ablation in ultrathin metallic foils [55]. Accordingly, this lowest frequency was set as the initial default condition in the Threshold Mapping Matrix (TMM), allowing independent optimization of peak power and pulse width while maintaining consistent average power delivery. As long as clean and continuous cuts were obtained under this setting, the frequency was kept constant and serving as a rule-of-thumb baseline for thin foil micromachining. Only in cases where penetration was incomplete, particularly for foils with higher thermal conductivity, the frequency increased incrementally (up to 10–15 kHz) as a single non-interactive adjustment to achieved edge continuity without modifying the overall matrix structure. Since the purpose of this study was to establish a rapid mapping optimization method rather than an exhaustive parametric analysis, frequency optimization was intentionally excluded from the multidimensional matrix and reserved for future work.

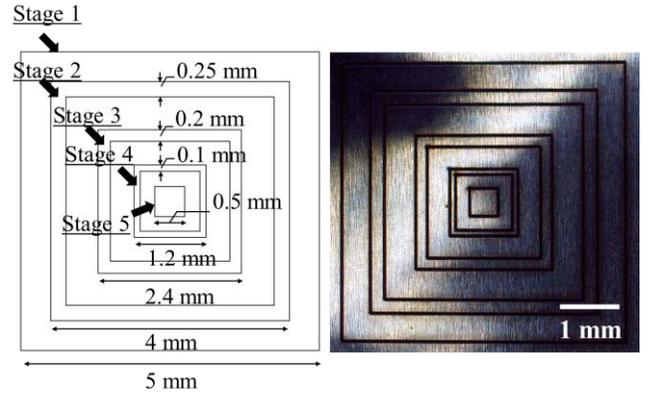

**Fig. 11.** Square-in-box verification pattern. Left: schematic with five stages of decreasing size and gap. Right: example of defect-free cut under optimal parameters.

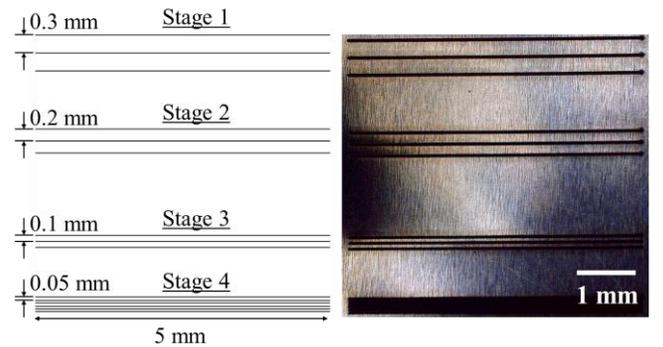

**Fig. 12.** Line array verification pattern. Left: schematic with four stages of decreasing gap from 0.50 mm to 0.05 mm. Right: defect-free lines under optimal parameters.

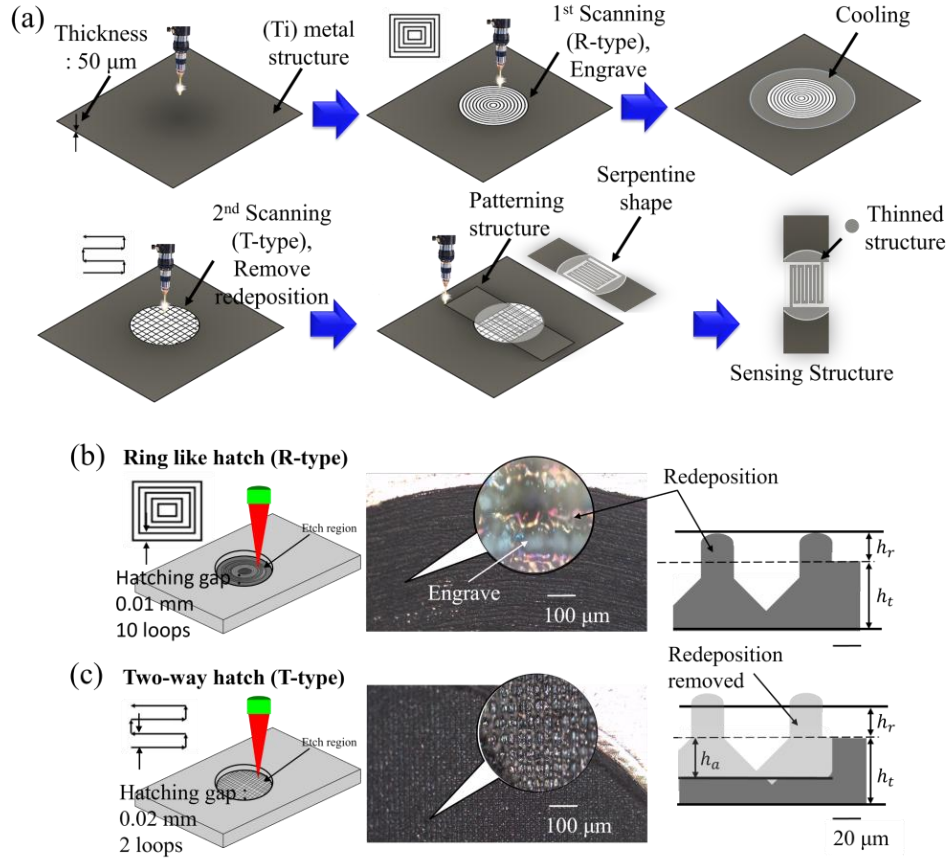

**Fig. 13.** Process flow for structure thinning: (a) overall process, (b) single-scan R-type, and (c) sequential-scan T-type.

Following TMM identification, an Energy Density Matrix (EDM) refined scanning dynamics and stabilized edge quality (Fig. 10). Laser power was fixed at the TMM-determined threshold, while scanning speed was varied from 3 to 80 mm/s. The optimal processing window was defined as the fastest scanning speed producing consistent, continuous cuts without excessive melting, burrs, or incomplete penetration.

By combining the TMM for penetration mapping and the EDM for scan-speed optimization, this dual-matrix approach provides a reproducible, data-driven method for selecting laser parameters for ultrathin metal foil micromachining, enabling stable fabrication of fine interdigitated structures.

### 3.3.3 Verification Patterns

To evaluate the effectiveness of the optimized parameters from the Threshold Mapping Matrix (TMM) and Energy Density Matrix (EDM), two verification patterns were implemented. These patterns cover different feature sizes and thermal load conditions.

**Verification Pattern 1:** The Square-in-box pattern (Fig. 11) consists of five concentric square stages having equal side length in both width and height. The layout is as follows: Stage 1, a 5.0 mm square; Stage 2, 4.0 mm and 3.5 mm squares separated by 0.25 mm; Stage 3, 2.4 mm and 2.0 mm with 0.20 mm spacing; Stage 4, 1.2 mm and 1.0 mm with 0.10 mm spacing; and Stage 5, a 0.50 mm square.

This configuration creates a thermal gradient from low confinement (Stage 1) to high confinement (Stage 5). If Stage 5, the highest confinement is cut cleanly with sharp edges, no visible burrs, and no discoloration, it implies the parameters are robust enough to handle a broad range of feature sizes. This makes the Square-in-box an efficient "worst-case scenario" test for intricate geometries.

**Verification Pattern 2:** The Line array pattern (Fig. 12) is designed to evaluate the process in high density linear cutting conditions, where parallel cuts may thermally interact. It contains four stages of parallel lines with progressively reduced spacing: Stage 1 at 0.50 mm, Stage 2 at 0.20 mm, Stage 3 at 0.10 mm, and Stage 4 at 0.05 mm.

The spacing values were chosen using a stepwise reduction toward the anticipated lower limit of the process resolution. At very small gaps, the heat-affected zones from adjacent cuts may overlap, leading to visible defects such as bridging, edge melting, or kerf widening. By including the 0.05 mm spacing in Stage 4, the pattern effectively tests whether the optimized parameters can maintain cut separation at the process limit. The clean cuts, parallel edges without thermal merging in Stage 4 strongly indicate that the parameter setup is at optimal conditions.

### 3.4 Selective Thinning Process

Fabricating thinned microstructures on titanium foil (50 µm) is challenging due to thermal effects. In preliminary trials, single pass ablation produced redeposited material that locally increased thickness rather than reducing it, as illustrated in Fig. 13(b). To overcome this, a sequential thinning method was established, as outlined in Fig. 13(a).

The process involved double scanning. The first step used an R-type circular hatching pattern (10 µm spacing, 10 loops) to initiate material removal. Following this pass, the foil was cooled in ethanol to suppress thermal accumulation. A sequential scanning with a T-type bidirectional hatching pattern (10 µm spacing, 2 loops) was then applied to remove redeposited debris and produce a cleaner ablation profile, as shown in Fig. 13(c). This combination countered the melt and redeposition effects observed in single pass ablation, enabling stable thinning of the foil.

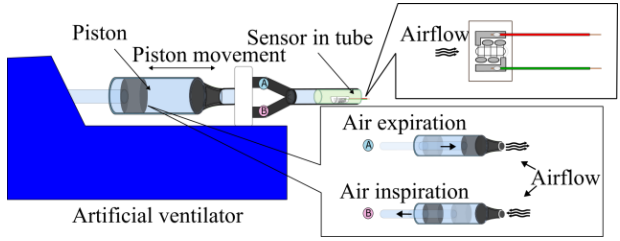

**Fig. 14.** Experimental setup for airflow demonstration using an artificial ventilator.

Thickness reduction in the ablated regions was confirmed using a digital micrometer (Mitutoyo, 1 μm resolution), and electrical resistance was measured after patterning the foil into a serpentine structure. The serpentine layout replaced the initial I-shaped geometry, which yielded insufficient resistance, and successfully extended the conductive path within the same footprint to achieve a target resistance of ~10 Ω. The final serpentine consisted of 16 turns, each 5 mm in length with a line width of 0.1 mm, giving a total conductor length of approximately 80 mm within a compact area, which provided adequate resistance for thermal sensing. Post fabrication measurements, however, indicated an actual line width of 0.06 ± 0.005 mm, consistently observed across multiple samples. This narrowing is attributed to the Gaussian beam profile and localized edge reflow during laser ablation, which tends to contract the final kerf relative to the CAD design.

To further validate the effect of sequential thinning, SEM images were captured after the first (R-type) and second (T-type) scans. These images confirmed that the cleaning step effectively reduced redeposition and improved surface uniformity. Taking together, these results establish that the selective thinning methodology provides a reliable early-stage strategy to reduce thermal mass and achieve adequate resistance in titanium sensing elements, thereby enabling functional MEMS flow sensor structures.

### 3.5 Experimental Setup for Airflow Sensing

The performance of the fabricated sensor was evaluated through a series of thermal and airflow experiments, beginning with the determination of the temperature coefficient of resistance (TCR). The sensing structure was placed in a temperature-controlled oven, with a thermocouple positioned 10 mm away to monitor ambient conditions. Resistance measurements were obtained using a Yokogawa Model 7562 digital multimeter, selected for its stability and noise immunity. The TCR was calculated from the slope of the resistance–temperature curve, providing a direct measure of the thermal sensitivity of the laser-thinned titanium element.

To further assess thermal behavior, surface temperature distribution was visualized using infrared thermography (InfReC R550Pro, Nippon Avionics Co., Ltd.). Two structures were examined under identical heating conditions: one bonded directly to a silicone rubber substrate and the other suspended across the cavity. Both were coated with black paint to standardize emissivity (0.94), ensuring reliable absolute temperature comparison.

Calibration of the flow sensor was performed by recording output voltages at known flow rates supplied using a mass flow controller (Model 3200, KOFLOC Corp.). From this, a calibration curve was constructed to correlate sensor output to volumetric flow. The recorded data was then plotted and approximated using King's law, which relates the heat loss from a heated wire to the fluid velocity. According to King's law, the relationship between the voltage, $V$ supplied to the sensor and the fluid velocity, $U$ can be expressed as in Eq. (7) where $A$, $B$, and n are empirically determined constants. By analyzing the data and generating a linear dependency graph, a calibration curve was established, accurately translating the sensor's voltage output into precise airflow measurements.

$$V^2 = A + BU^n \qquad (7)$$

Finally, sensor performance under physiological conditions was tested using a commercial artificial ventilator (Small Animal Ventilator Model 683, Harvard Apparatus), as illustrated in Fig. 14. The ventilator generated both sinusoidal and pulse-modulated airflow profiles with adjustable tidal volumes (0.5–1.5 L) and flow rates (0–30 L/min). The sensor was installed in a custom acrylic chamber connected to the ventilator tubing. A constant current source supplied the bias current for Joule heating, while voltage signals across the sensor were recorded using a data acquisition system. As airflow increased, the cooling effect reduced the temperature of the heated sensing element, leading to a corresponding decrease in resistance and output voltage. The voltage-time response was analyzed to assess the thermal behavior and responsiveness of the sensor.

## 4. Result and Discussion

### 4.1 Structural Optimization using TMM - EDM

The TMM experiments established clear penetration thresholds for ultrathin foils. For titanium at 50 μm, stable cutting was achieved at peak power 14 W with pulse widths of 150–200 ns at 5 kHz, corresponding to ~10.5 mW average power. Reducing the thickness to 20 μm lowered the threshold to peak power 6 W, giving ~4.5 mW. Nichrome at 20 μm required peak power 10 W at similar pulse widths giving ~7.5 mW. In contrast, copper, owing to its higher thermal conductivity, demanded both peak power 10 W and an elevated frequency of 35 kHz to confine localized heating and avoid incomplete penetration.

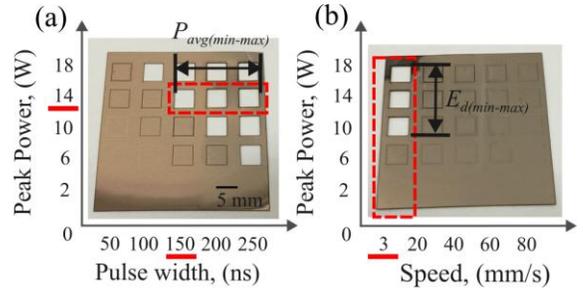

**Fig. 15.** Experimental cut results for titanium (50 μm) using the (a) TMM and (b) EDM approach.

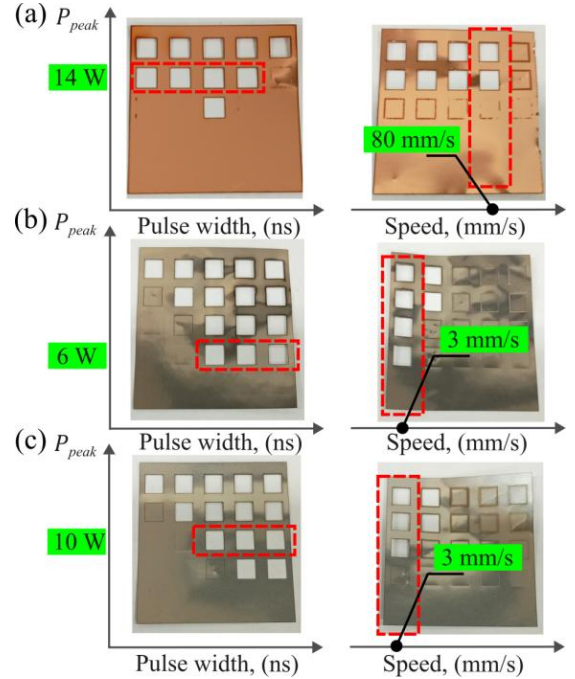

**Fig. 16.** Cutting analysis on: (a) copper (50 μm), (b) titanium (20 μm), and (c) nichrome (20 μm).

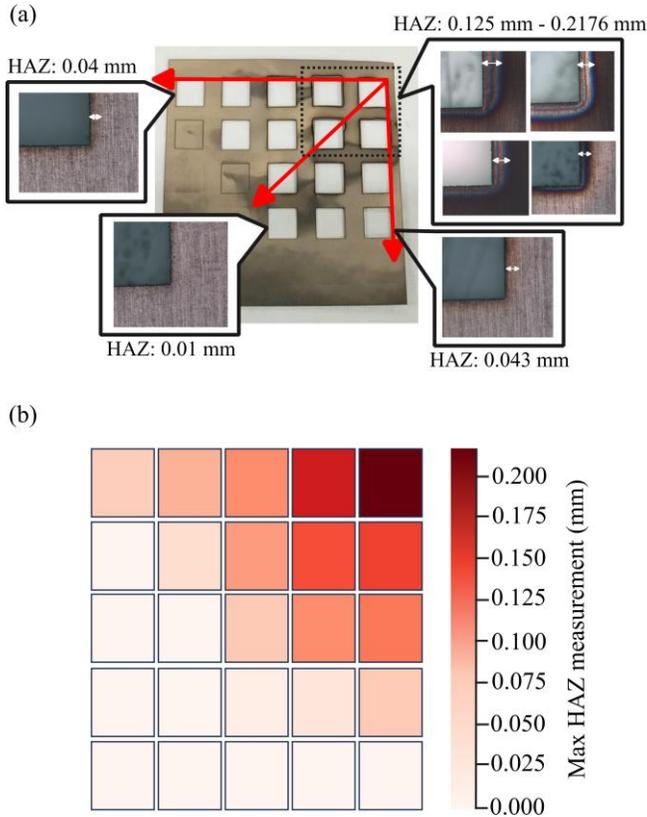

(a)

HAZ: 0.04 mm

HAZ: 0.125 mm - 0.2176 mm

HAZ: 0.01 mm

HAZ: 0.043 mm

(b)

**Fig. 17.** HAZ observation on TMM-EDM optimization; (a) Photograph, (b) 2D heatmap of HAZ measurement.

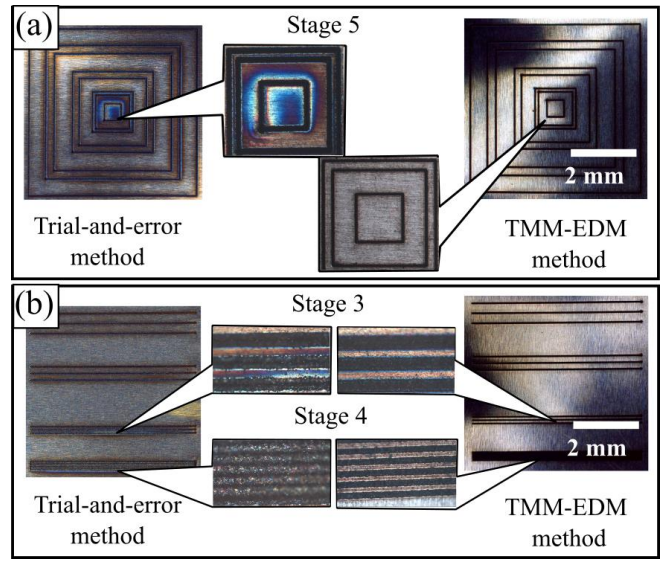

**Fig. 18.** A photograph of trial-and-error method and propose TMM-EDM method: (a) Square-in-box verification pattern, (b) Line array verification pattern.

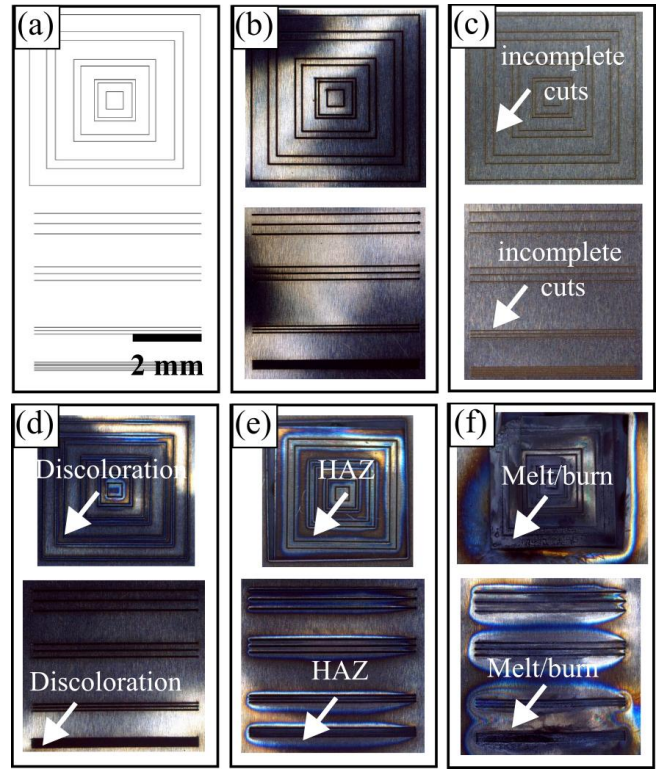

**Fig. 19.** A photograph of: (a) Validation patterns, (b) clean cut quality and (c)-(f) defect conditions at all stage.

After penetration thresholds were mapped, EDM refinement was carried out to stabilize cut quality through control of scanning speed. For titanium and nichrome foils, the optimal window was consistently found at 3 mm/s, producing sharp kerfs with minimal heat-affected zones (HAZ). Copper, however, required a much higher scan speed of 80 mm/s to balance its rapid thermal spreading.

The results highlight that threshold values scale not only with thickness but also with thermal conductivity, a parameter that strongly governs heat confinement during laser–material interaction. For low-conductivity metals such as titanium and nichrome, slow scanning at moderate power produced stable penetration without excessive melt. In contrast, for high-conductivity copper, localized heating could only be maintained through higher repetition rates and rapid scanning. This material-dependent behavior confirms that the framework is not restricted to a single case but can be systematically extended to metals of different properties by tuning the same set of variables (peak power, pulse width, frequency, and scan speed). The demonstrated consistency across three metals of varying thickness establishes the method's transferability and provides a reproducible pathway for micromachining ultrathin foils with diverse thermal characteristics.

The resulting HAZ behavior, as observed in the sample analysis (Fig. 17), directly reflects this optimization sequence. The largest HAZ dimensions were observed in the top right region of the mapped, corresponding to the initial or less refined cutting parameters tested during the TMM-EDM mapping. Conversely, the minimal HAZ values (as low 0.01 mm) were found in the areas corresponding to the optimized process window defined.

### 4.2 Verification Pattern Analysis

The optimized parameters were validated using two distinct test patterns: a square-in-box structure and a line array.

Figure 18 presents a direct comparison between the proposed dual-matrix optimization and the conventional trial-and-error approach, highlighting the significant advantages of the TMM–EDM framework. Under optimized TMM–EDM conditions, both patterns exhibited defect-free cuts, with no observable HAZ, discoloration, or burr formation across all confinement levels. Even at the smallest line spacing of 50 µm, the line array maintained parallel integrity without thermal bridging.

Table 1.  Parameters for laser cutting (Titanium (TC: 21.9 Wm⁻¹K⁻¹), Copper (TC: 401 Wm⁻¹K⁻¹) and Nichrome (TC: 11.3 Wm⁻¹K⁻¹) [54].

| Material | TMM | | EDM | |
|---|---|---|---|---|
| | $P_{(T.avg)}$ (W) | $\tau_T$ (ns) | $f_T$ (kHz) | $v$, (mm/s) |
| Titanium (50 µm) | 14 | 150 | 5 | 3 |
| Copper (50 µm) | 14 | 150 | 35 | 80 |
| Titanium (20 µm) | 6 | 150 | 5 | 3 |
| Nichrome (20 µm) | 10 | 150 | 5 | 3 |

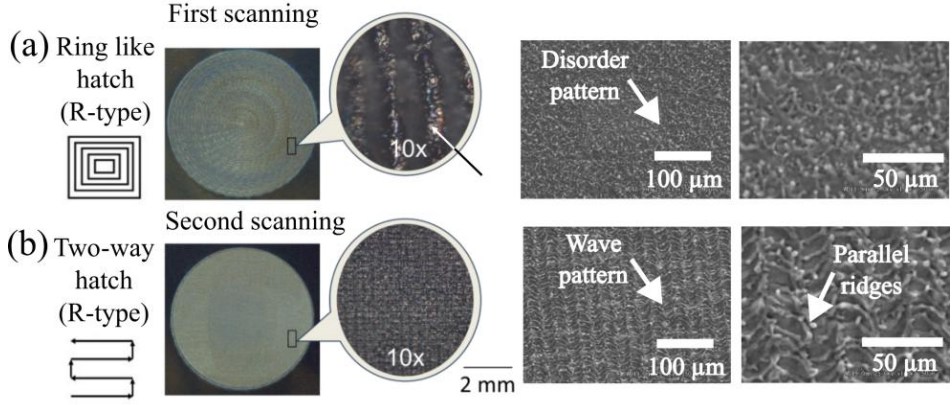

**Fig. 20.** Photographs and SEM images of the interdigitated device structure: (a) thinning region after single scanning, (b) thinning region after sequential scanning.

In the previous work [55], parameter optimization was performed sequentially by varying one factor at a time such as laser power, scanning speed, or pulse frequency while keeping the remaining parameters constant. For example, laser power was adjusted from 8 to 14 W, scanning speed from 50 to 5 mm/s, and frequency from 50 to 5 kHz, each requiring multiple experimental iterations to identify the optimum condition. In total, roughly 20 parameter combinations were required to compare and obtain acceptable cutting quality, but the process was time consuming and demanded extensive experimentation.

In contrast, the proposed TMM–EDM framework enables simultaneous evaluation of multiple parameter interactions within a single experimental sequence. Using the matrix based mapping strategy, the fiber laser control software assigns distinct combinations of power, pulse width, and scanning speed to predefined regions of the foil in one cutting operation. After a single processing run, the resulting pattern provides a comprehensive visual map from which the optimal cutting window can be directly identified based on continuous and clean cuts. If clean penetration is not achieved on first trial, second evaluation is performed by slightly increasing the pulse repetition frequency, while maintaining the matrix structure. This approach drastically reduces the number of experimental iterations and consistently yields uniform, burr-free edges across all confinement levels.

**Table 2.** Surface roughness produced by three processing strategies.

| Parameter | Single-scan R-type | Sequential-scan T-type | Double scan (2 loops) |
|---|---|---|---|
| Ra (nm) | 4818.55 | 3503.51 | 2486.43 |
| Rq (nm) | 6389.46 | 4763.32 | 3068.09 |
| Rt (nm) | 40397.98 | 28218.55 | 16665.66 |
| Rz (nm) | 40388.76 | 28204.35 | 16660.54 |
| Rsk | 0.08 | 0.11 | 0.07 |
| Rku | 3.83 | 3.61 | 2.77 |

When parameters deviated from the optimized window, defects became pronounced. As shown in Fig. 19 (c–f), incomplete penetration, edge roughness, and localized melting occurred, particularly at higher confinement levels. These defects are consistent with excess thermal load and uneven heat distribution, reinforcing that both penetration and scan stability must be jointly optimized.

Table 1 summarizes the optimized parameters for each tested material. These values define practical operating ranges that can be directly applied to fabrication of sensing structures, providing a repeatable alternative to ad hoc optimization and ensuring clean-cut quality suitable for downstream MEMS integration.

### 4.3 Thinning process performance (sequential R-T scans)

The purpose of the thinning experiment was to validate whether sequential R–T laser scans could reliably reduce the thickness of Ti foils while maintaining structural stability, thereby achieving the resistance levels required for thermal flow sensing in free-standing structures. Initial validation was performed on serpentine-shaped test structures before transferring the same approach to the interdigitated sensing geometry. Fig. 20 shows an optical observation of first and sequential second thinning scanning. A single R-type scan produced visible redeposition and surface irregularities, sometimes giving the appearance of thickening rather than thinning. When a secondary T-type scan was added, the redeposited layer was effectively removed, resulting in a smoother surface.

The surface morphology differences between first and sequential second scans are highlighted in the SEM images of Fig. 20 (a-b). The single R-type scan produced rough, irregular surfaces with visible redeposition and localized melting. In contrast, the sequential R–T scans yielded uniform periodic textures and cleanly ablated regions with minimal residual debris.

To further quantify the surface effects observed in SEM, quantitative surface profiling was also performed using a Bruker Dektak XT-E stylus profilometer (vertical range: 10 nm–1 mm, vertical resolution up to 0.1 nm), as shown on Fig. 21 and summarized in Table 2. At First, the single R-type scan was compared with a double-scan approach, in which a secondary T-type scan was applied immediately after the first scan. Then, applying the R-type + T-type sequence for two loops was also evaluated as a further continue process increase etching depth. Quantitative results indicate a clear trend of surface smoothing with successive processing. Ra decreased from 4818 nm (single R-type scan) to 3504 nm (sequential-scan T-type) and further to 2486 nm (R-type + T-type sequence for two loops), representing a reduction of 48% relative to the single scan. Rq, Rt, and Rz follow a similar trend, confirming attenuation of both average deviations and peak-to-valley extremes. The height distribution histogram (Fig. 21) demonstrates narrowing of distribution with successive processing, consistent with Ra/Rq reductions. In this histogram, the 'Count' represents the number of sampled data points measured at a specific height deviation (μm) from the mean line. Skewness (Rsk) remains small and positive for all cases, indicating slight dominance of peaks over valleys, while kurtosis (Rku) decreases from ~3.83 (single R-type scan) to ~2.77 (R-type + T-type sequence for two loops), indicating the distribution becomes less heavy-tailed and more platykurtic after the 2-loop process. This combination of SEM imaging and profilometry confirms that the sequential R–T process not only removes redeposited material but also systematically suppresses surface asperities as the thinning depth increases.

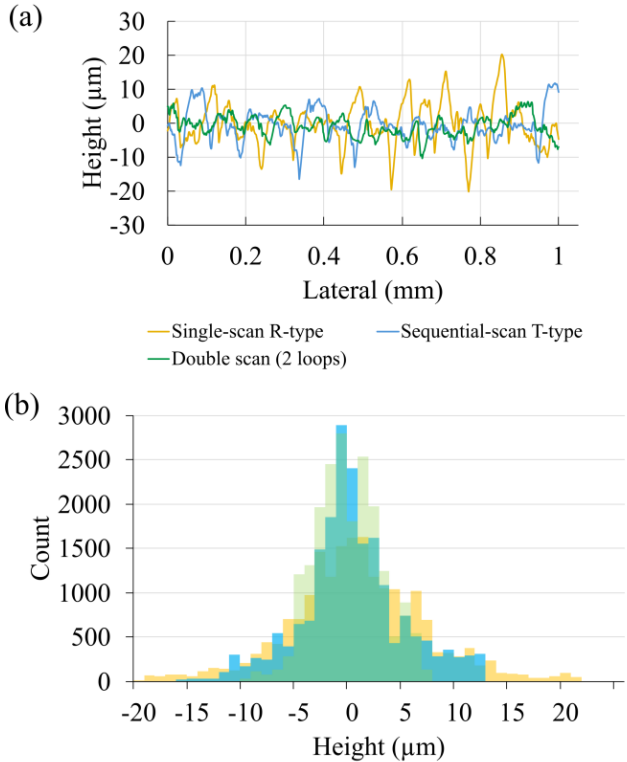

**Fig. 21.** Surfaces roughness profiles; (a) Surface profile comparison, and (b) Height distribution comparison.

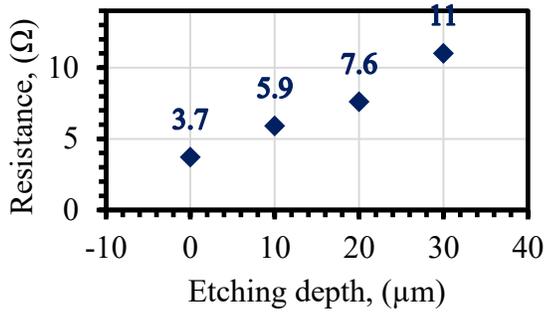

**Fig. 22.** Resistance of the final sensing structure versus etching depth from the thinning process.

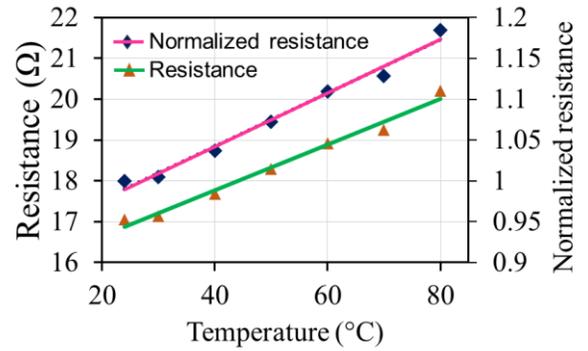

**Fig. 24.** Temperature dependence of titanium film resistance.

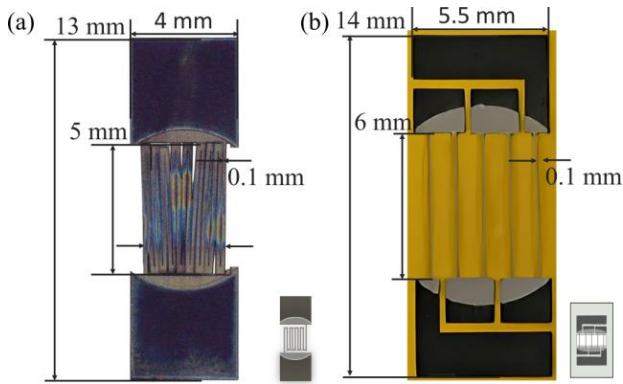

**Fig. 23.** Photographs of the Ti foil laser thinning process: (a) Serpentine structure and (b) Interdigitate structure.

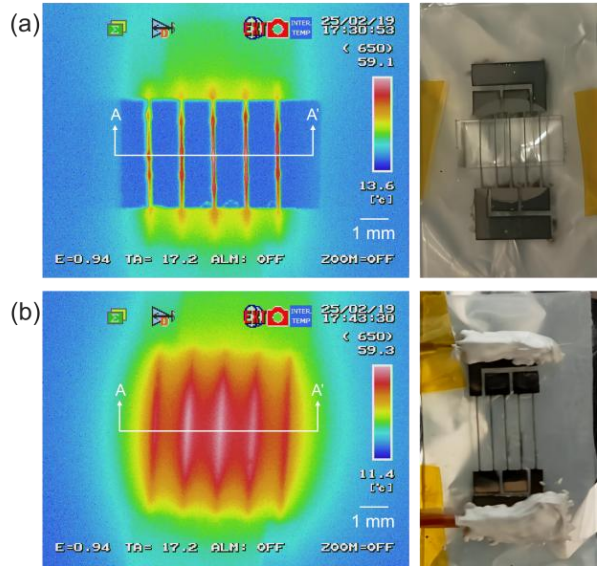

**Fig. 25.** Thermography images of (a) free-standing structure (with cavity) and (b) structure fixed with substrate (without cavity).

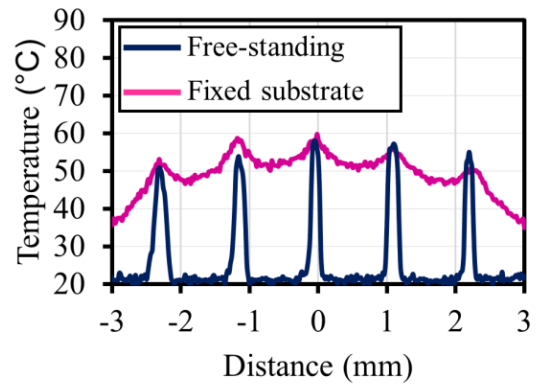

**Fig. 26.** Temperature profile of the A-A' as indicated in Fig. 25.

The electrical impact of thinning the serpentine structure is summarized in Fig. 22. Without thinning, the baseline resistance of the 50 μm foil was 3.7 Ω. After one thinning cycle (~10 μm depth), resistance increased to 5.9 Ω. A second cycle (~20 μm depth) yielded 7.6 Ω, and a third cycle (~30 μm depth) reached 11 Ω, which corresponds to the design target for constant-temperature operation. This clear correlation between etching depth and resistance demonstrates that the sequential R–T scans not only reduce thickness but also achieves the resistance necessary for sensor operation. However, attempts beyond three cycles caused tearing of the foil, indicating that ~30 μm thinning represents the practical structural limit for stability. It should be noted that the tearing observed when the foil thickness was reduced below 30 μm was not due to mechanical failure but instead caused by excessive thermal energy during laser ablation. At such thin residual layers, localized plasma pressure and heat accumulation can destabilize the molten boundary, producing microvoids and tearing along the ablation path. This effect represents the thermal process limit of the thinning mechanism.

Earlier prototypes using a serpentine layout provided a higher nominal resistance but exhibited instability during airflow testing, as the unsupported foil section oscillated under flow excitation, resulting in fluctuating resistance. For this reason, the final design adopted an interdigitated configuration with an integrated supporting frame, ensuring a stable free-standing region that preserves consistent resistance and response characteristics even when mounted along a curved tube or subjected to airflow disturbances.

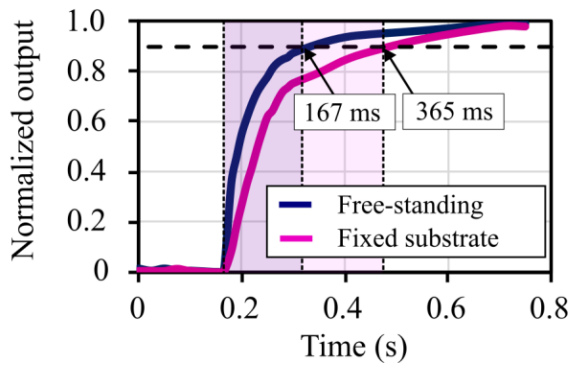

**Fig. 27.** Response to a step input of each structure.

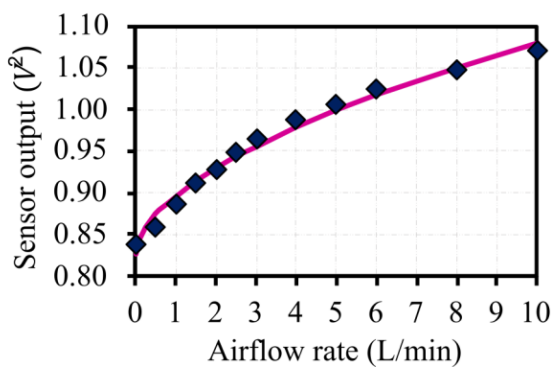

**Fig. 28.** Calibration curve of the fabricated sensor with free-standing structure.

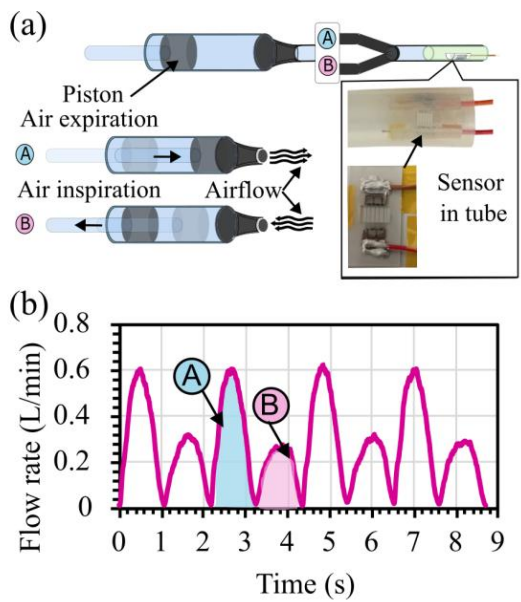

**Fig. 29.** Airflow rate sensing demonstration; (a) Experimental setup by using artificial ventilator, and (b) Flow rate measurement by conversion sensor.

After resolving the mechanical stability issue, the same thinning procedure was applied to the finalized interdigitated sensing element (Fig. 23). The unthinned structure exhibited an initial resistance of approximately 4–5 Ω, which was insufficient for thermal flow operation. Following sequential thinning, the resistance increased to ~11 Ω. This confirms that once the thinning parameters are established, the dual-scan strategy can be reliably transferred to more complex geometries without additional iteration, achieving both the required electrical characteristics and maintaining structural integrity.

Overall, the results from Figs. 20–23 confirm that the sequential thinning process enables stable, firm interdigitated structures with the required resistance range, overcoming redeposition and deformation issues inherent to single-pass ablation, and providing a reliable fabrication pathway toward MEMS-compatible titanium flow sensors.

### 4.4 Thermal Response Analysis

The temperature coefficient of resistance (TCR) was first measured to evaluate the sensitivity of the selectively thinned titanium sensing structure. As shown on Fig. 24, the resistance increased linearly with temperature, yielding a TCR of 3278 ppm/°C at 25 °C. This result is consistent with reported values for titanium microheaters, such as the 4146 ppm/°C obtained in a recent comparative study [56]. The R-squared value for this measurement was 0.9843, indicating a strong linear correlation between resistance and temperature.

Thermal imaging was conducted to evaluate the thermal insulation advantage of the free-standing configuration compared to a substrate-fixed structure. The sensing elements were actuated at 60 °C using the feedback-controlled circuit shown in Fig. 3, and their surface temperature distributions were recorded under steady-state operation. The results are shown in Fig. 25 for the two structural configurations. The free-standing structure with a cavity (Fig. 25a) exhibited concentrated heating along the suspended wires, with the thermal distribution confined to narrow regions that closely follow the geometry of the sensing elements. This localization arises because the cavity reduces heat conduction paths, limiting dissipation into the surrounding substrate. By contrast, the substrate-fixed structure (Fig. 25b) showed a much wider heat distribution, as the underlying silicone acted as a thermal sink, spreading the heat laterally and reducing the temperature contrast across the wires.

The extracted temperature profiles (Fig. 26) reinforce these observations with quantitative evidence. For the free-standing design, the blue curve shows five distinct peaks, each aligned with a sensing wire, reaching ~55–60 °C while the baseline between wires returns to near ambient (~20–25 °C). This sharp contrast demonstrates that heating is confined almost exclusively to the active elements. In comparison, the fixed-substrate design (pink curve) shows a broad thermal plateau, maintaining ~40–45 °C even between wires, and peak temperatures only marginally higher (~50 °C). The broader distribution confirms significant lateral heat leakage into the substrate. The difference of nearly 30 °C between wire and baseline in the free-standing case, compared to less than 10 °C in the fixed case, highlights the superior localization and reduced thermal mass achieved by suspension. This confinement directly translates to lower power requirements and faster thermal response, both of which are critical advantages for airflow sensing applications.

### 4.5 Airflow Sensing Performance

The dynamic response of the fabricated sensor was first evaluated under step-input excitation. A solenoid valve was installed in between the mass flow controller airflow supply and the sensor for switching. Both the free-standing structure and the sensor installed on the substrate was subjected to the test. The free-standing structure exhibited a significantly faster thermal response than the substrate-fixed design. As shown in Fig. 27, the response time improved from 365 ms to 167 ms, representing a 54% reduction. This enhancement is directly attributed to the reduced thermal mass and minimized heat leakage enabled by the suspended configuration, consistent with the localized heating observed in the thermography results.

To assess steady-state sensing performance, airflow rate sensing calibration was performed under controlled environmental conditions (12.9 °C, 43.9% relative humidity) to ensure data consistency and repeatability. For the practical breathing demonstration, temperature and humidity differences may slightly influence baseline resistance, however, the dominant trend in resistance changes due to airflow remained consistent with the calibration response. Future work will include calibrated respiratory testing under controlled ambient conditions to quantitatively correlate breathing flow rates with the calibrated sensitivity curve. The sensor's output exhibited a clear adherence to King's Law, with the calibration curve achieving a coefficient of determination ($R^2$) of 0.986, indicating high measurement accuracy as depicted in Fig. 28. The relationship between the sensor output voltage and the airflow rate is expressed by the calibration equation given in Eq. (8);

$$V^2 = 0.82709 + 0.0698 U^{0.56} \qquad (8)$$

During this calibration, the electrical power required to maintain the sensing wire under constant-temperature operation was recorded. The device consumed approximately 50 mW, corresponding to the steady heating of the thinned Ti element. While continuous operation was used for characterization in this study, the airflow in practical respiratory monitoring does not require uninterrupted sampling. Therefore, the heater can be driven intermittently using a pulse-width-modulated (PWM) excitation scheme, which would substantially reduce the average power consumption. Such duty-cycled operation will be explored in future work to adapt the sensor toward portable and wearable biomedical platforms.

The sensor's applicability to practical respiratory monitoring was further validated using an artificial ventilator (Model 683, Harvard Apparatus) operating at 0.5 Hz. As shown in Fig. 29 (a), the sensor was inserted into the connector tube with an inner diameter of 6.5 mm and fixed to the tube wall for measurement. Airflow was generated by a mechanical cylinder, with inspiration and expiration phases controlled via a three-port valve. Since the calibration was conducted for one-directional flow, the integration of the flow rate over time, as shown in Fig. 29 (b), was calculated for the air expiration phase （A）only. The results reveal that the ventilator delivers approximately 26 cm³ per cycle during air inspiration, compared to the nominal piston stroke of 30 cm³, with the ~13% deviation attributed to tubing compliance and mechanical tolerances of the analog ventilator rather than sensor inaccuracy. Importantly, the sensor consistently resolved tidal volumes within the clinically relevant range, underscoring its suitability for respiratory monitoring applications.

The response time, calibration, and ventilator integration results demonstrate that the developed free-standing titanium sensor provides both high temporal resolution and reliable quantitative accuracy, addressing the key requirements for biomedical airflow sensing.

This study primarily focuses on the fabrication feasibility and airflow sensing performance of the proposed single-step laser patterning and thinning titanium flow sensor. The initial characterization demonstrates stable response and repeatability during airflow evaluation, establishing the concept's potential for compact and biocompatible sensing structures. Although long-term reliability was not the central scope of this work, it remains a significant consideration for future device deployment. Nevertheless, existing literature strongly supports the stability of titanium-based materials in biomedical environments. Well-polished Ti and Ti–6Al–4V alloys have shown exceptional fatigue endurance exceeding $10^7$ cycles when surface defects are controlled [57][58]. Based on these findings and the consistent behavior observed during repeated experimental use, the current free-standing structure is considered mechanically and chemically stable under typical operating conditions.

*4.6 Discussion of Biocompatibility and Application Scope*

The use of titanium as the primary sensing material provides inherent advantages in biological environments. From a fabrication standpoint, the demonstrated laser-based thinning and structuring process avoids wet chemical etching, offering a cleaner and safer route to biocompatible sensor production. The selective thinning approach also enables resistance tuning without introducing foreign coatings or alloying, thereby preserving material purity. SEM observations confirmed that sequential scanning effectively minimized redeposition and surface irregularities, ensuring structural stability for long-term biocompatibility.

In terms of application scope, the reduced thermal mass and improved response time achieved by the free-standing design directly translate to higher sensitivity for respiratory flow monitoring, as demonstrated in ventilator testing. The combination of high TCR, accurate airflow calibration, and stable volume measurement supports its potential for integration into medical devices such as ventilators, Continuous Positive Airway Pressure (CPAP) machines, or wearable respiratory monitors. Furthermore, the scalability of the dual-matrix optimization framework allows extension to other ultrathin biocompatible metals, broadening applicability to implantable thermal flow or pressure sensors.

## 5. Conclusion

This work demonstrated a single-step laser-based fabrication strategy for MEMS thermal flow sensors using biocompatible metallic foils. The process integrates patterning and localized thinning within a unified fiber laser micromachining platform, eliminating the need for multi-step photolithographic and chemical processing. The dual-matrix optimization framework, consisting of the Threshold Mapping Matrix (TMM) and Energy Density Matrix (EDM), enabled systematic determination of penetration thresholds and scanning parameters to achieve defect-free micro cutting. A sequential R–T scanning strategy was introduced to address redeposition effects associated with the Gaussian beam profile. This method achieved uniform thickness reduction from 50 μm to 20–30 μm while preserving the mechanical integrity of narrow interdigitated lines. The resulting free-standing, locally thinned structures exhibited significantly reduced thermal mass and improved transient response. Airflow sensing experiments confirmed accurate velocity calibration with ($R^2$) = 0.986, a 54% improvement in response time compared to substrate-supported designs, and reliable performance in cyclic flow conditions representative of respiratory monitoring. These results demonstrate that the proposed fabrication method is a scalable, material-independent strategy for realizing high-performance MEMS flow sensors and flexible, biocompatible microsystems.

## 6. Abbreviations

The following abbreviations are used in this manuscript: MEMS (Micro-Electro-Mechanical Systems), CTA (Constant-Temperature Anemometer/Anemometry), TCR (Temperature Coefficient of Resistance), RTD (Resistance Temperature Detector), TMM (Threshold Mapping Matrix), EDM (Energy Density Matrix), HAZ (Heat-Affected Zone), Nd:YAG (Neodymium-doped Yttrium Aluminum Garnet), UV (Ultraviolet), CO₂ (Carbon Dioxide, laser source), SEM (Scanning Electron Microscope), IEC (International Electrotechnical Commission), CAGR (Compound Annual Growth Rate), CPAP (Continuous Positive Airway Pressure), MEXT (Ministry of Education, Culture, Sports, Science and Technology, Japan), JSPS (Japan Society for the Promotion of Science), and KAKENHI (Grants-in-Aid for Scientific Research, Japan).

## 7. Competing Interests

The authors declare that they have no competing interests.

## 8. Consent for publication

Not applicable.

## 9. Ethics approval and consent to participate

Not applicable.

## 10. Funding

This work was partially supported by the Japan Society for the Promotion of Science (JSPS) KAKENHI Grant-in-Aid for Early-Career Scientists under Grant 24K21094, and the Advanced Research Infrastructure for Materials and Nanotechnology in Japan (ARIM) of the Ministry of Education, Culture, Sports, Science and Technology (MEXT) project no. JPMXP1225RO0017. The first author was funded by a scholarship from the Ministry of Education, Culture, Sports, Science, and Technology (MEXT) of Japan.

## 11. Availability of data and materials

Data will be made available on request.

## 12. Authors' contributions

All authors have read and approved the final manuscript.

## Acknowledgements

None.

## References

[1] Mordor Intelligence, MEMS Sensor Market Size & Share Analysis – Growth Trends & Forecasts (2025–2030), 2025. [Online]. Available: https://www.mordorintelligence.com/industry-reports/mems-sensor-market.

[2] Mordor Intelligence. (2025). Flow sensors market size & share analysis – growth trends & forecasts (2025–2030). https://www.mordorintelligence.com/industry-reports/flow-sensor-market.

[3] Mohd Javaid, Abid Haleem, Ravi Pratap Singh, Shanay Rab, Rajiv Suman, Significance of sensors for industry 4.0: Roles, capabilities, and applications, Sensors International, Volume 2, 100110, https://doi.org/10.1016/j.sintl.2021.100110.

[4] Mohammed Aarif K. O., Afroj Alam, Yousuf Hotak, Smart Sensor Technologies Shaping the Future of Precision Agriculture: Recent Advances and Future Outlooks, Journal of Sensors 2025, https://doi.org/10.1155/js/2460098.

[5] Vo, D.-K.; Trinh, K.T.L., Advances in Wearable Biosensors for Healthcare: Current Trends, Applications, and Future Perspectives. Biosensors 2024, 14, 560. https://doi.org/10.3390/bios14110560.

[6] Li, A.; Zhao, H.; Zhou, Y.; Liu,Z.. A Review of CMOS-MEMS Thermal Flow Sensor. Applied and Computational Engineering,168,87-98 (2025) https://doi.org/10.54254/2755-2721/2025.24253.

[7] Kang, W., Choi, HM. & Choi, YM. Development of MEMS-based thermal mass flow sensors for high sensitivity and wide flow rate range. J Mech Sci Technol 32, 4237–4243 (2018). https://doi.org/10.1007/s12206-018-0822-4.

[8] Balakrishnan, V.; Phan, H.-P.; Dinh, T.; Dao, D.V.; Nguyen, N.-T. Thermal Flow Sensors for Harsh Environments. Sensors 2017, 17, 2061. https://doi.org/10.3390/s17092061.

[9] Nasirzadeh, Z.; Ghasemi, M.M.; Fathi, A.; Tavakkoli, H. A Dual-Region MEMS Thermal Flow Sensor with Obstacle-Enhanced Sensitivity and Linearity Across Wide Velocity Ranges. Electronics 2025, 14, 2128. https://doi.org/10.3390/electronics14112128.

[10] F. Hedrich, K. Kliche, M. Storz, S. Billat, M. Ashauer, R. Zengerle, Thermal flow sensors for MEMS spirometric devices, Sensors and Actuators A: Physical, Volume 162, Issue 2, Pages 373-378, (2010) https://doi.org/10.1016/j.sna.2010.03.019.

[11] Judy, J. W., "Microelectromechanical systems (MEMS): Fabrication, design and applications", Smart Materials and Structures, vol. 10, pp. 1115-1134, 2001. https://doi.org/10.1088/0964-1726/10/6/301

[12] Tadjgadapa, S. A., and Najafi, N. (November 11, 2003). "Developments in Microelectromechanical Systems (MEMS): A Manufacturing Perspective ." ASME. J. Manuf. Sci. Eng. November 2003; 125(4): 816–823. https://doi.org/10.1115/1.1617286.

[13] Karimi, K.; Fardoost, A.; Mhatre, N.; Rajan, J.; Boisvert, D.; Javanmard, M. A Thorough Review of Emerging Technologies in Micro- and Nanochannel Fabrication: Limitations, Applications, and Comparison. Micromachines 2024, 15, 1274. https://doi.org/10.3390/mi15101274.

[14] M. Gower, "Laser micromachining for manufacturing MEMS devices," Proceedings of SPIE, vol. 4559, 2001 https://doi.org/10.1117/12.443040.

[15] Eckhard Beyer, Achim Mahrle, Matthias Lütke, Jens Standfuss, Frank Brückner, Innovations in high power fiber laser applications, Proceedings Volume 8237, Fiber Lasers IX: Technology, Systems, and Applications; 823717 (2012) https://doi.org/10.1117/12.910899.

[16] Bill Shiner, Fiber lasers for manufacturing, Proceedings Volume 5706, Critical Review: Industrial Lasers and Applications; (2005) https://doi.org/10.1117/12.601648.

[17] M. N. Mohamed Zukri, M. S. Al Farisi, Y. Hasegawa, M. Shikida, Laser micromachined 3D-shaped free-standing structures on flexible substrate for thermal flow sensor The 23rd International Conference on Solid-State Sensors Actuators and Microsystems (Transducers), Orlando, USA, 29 June – 3 July 2025, pp. 2125–2128.

[18] Dejene, N.D., Lemu, H.G. & Gutema, E.M. Effects of process parameters on the surface characteristics of laser powder bed fusion printed parts: machine learning predictions with random forest and support vector regression. Int J Adv Manuf Technol 133, 5611–5625 (2024). https://doi.org/10.1007/s00170-024-14087-5.

[19] Aakif Anjum, A.A. Shaikh, Nilesh Tiwari. Experimental investigations and modeling for multi-pass laser micro-milling by soft computing-physics informed machine learning using CO2 laser. Optics & Laser Technology, Volume 158, Part A, (2023) https://doi.org/10.1016/j.optlastec.2022.108922.

[20] Zhang, X., Zhou, L., Feng, G. et al. Laser technologies in manufacturing functional materials and applications of machine learning assisted design and fabrication. Adv Compos Hybrid Mater 8, 76 (2025). https://doi.org/10.1007/s42114-024-01154-4.

[21] Nur Islahudin, Dony Satriyo Nugroho, et al Machine learning-driven optimization for surface roughness prediction of vertical orientation measurements on 3D printed components. Cleaner Engineering and Technology, Volume 28 (2025). https://doi.org/j.clet.2025.101046.

[22] Cevik, Z.A.; Ozsoy, K.; Ercetin, A.; Sariisik, G. Machine Learning-Driven Optimization of Machining Parameters Optimization for Cutting Forces and Surface Roughness in Micro-Milling of AlSi10Mg Produced by Powder Bed Fusion Additive Manufacturing. Appl. Sci. 2025, 15, 6553. https://doi.org/10.3390/app15126553.

[23] Zoheir Kordrostami, Kourosh Hassanli, Amir Akbarian; MEMS piezoresistive pressure sensor with patterned thinning of diaphragm. Microelectronics International 11 June 2020; 37 (3): 147–153. https://doi.org/10.1108/MI-09-2019-0060.

[24] Shubham, S.; Seo, Y.; Naderyan, V.; Song, X.; Frank, A.J.; Johnson, J.T.M.G.; da Silva, M.; Pedersen, M. A Novel MEMS Capacitive Microphone with Semiconstrained Diaphragm Supported with Center and Peripheral Backplate Protrusions. Micromachines 2022, 13, 22. https://doi.org/10.3390/mi13010022.

[25] Han, X., Huang, M., Wu, Z. et al. Advances in high-performance MEMS pressure sensors: design, fabrication, and packaging. Microsyst Nanoeng 9, 156 (2023). https://doi.org/10.1038/s41378-023-00620-1.

[26] Song P, Ma Z, Ma J, Yang L, Wei J, Zhao Y, Zhang M, Yang F, Wang X. Recent Progress of Miniature MEMS Pressure Sensors. Micromachines (Basel). 2020 Jan 1;11(1):56. doi: 10.3390/mi11010056.

[27] M. S. Al Farisi, Y. Wang, Y. Hasegawa, M. Matsushima, T. Kawabe and M. Shikida, "Facile In-Tube-Center Packaging of Flexible Airflow Rate Microsensor for Simultaneous Respiration and Heartbeat Measurement," in IEEE Sensors Journal, vol. 23, no. 12, pp. 12626-12633, 15 June15, 2023, doi: 10.1109/JSEN.2023.3272310.

[28] Washim Reza Ali, Mahanth Prasad, Fabrication of microchannel and diaphragm for a MEMS acoustic sensor using wet etching technique, Microelectronic Engineering, Volume 253, (2022) https://doi.org/10.1016/j.mee.2021.111670.

[29] Algamili, A.S., Khir, M.H.M., Dennis, J.O. et al. A Review of Actuation and Sensing Mechanisms in MEMS-Based Sensor Devices. Nanoscale Res Lett 16, 16 (2021). https://doi.org/10.1186/s11671-021-03481-7.

[30] Li, Chaojiang, Yang, Yuxin, Qu, Rui et al. Recent advances in plasma etching for micro and nano fabrication of silicon-based materials: a review. Journal of Materials Chemistry C, 2024,12, 18211-18237. https://doi.org/10.1039/D4TC00612G.

[31] Huff, M. Recent Advances in Reactive Ion Etching and Applications of High-Aspect-Ratio Microfabrication. Micromachines 2021, 12, 991. https://doi.org/10.3390/mi12080991.

[32] Michael Bauhuber, Andreas Mikrievskij, Alfred Lechner. Isotropic wet chemical etching of deep channels with optical surface quality in silicon with HNA based etching solutions, Materials Science in Semiconductor Processing, Volume 16, Issue 6, Pages 1428-1433, 2013. https://doi.org/10.1016/j.mssp.2013.05.017.

[33] Kirsch, B.; Bohley, M.; Arrabiyeh, P.A.; Aurich, J.C. Application of Ultra-Small Micro Grinding and Micro Milling Tools: Possibilities and Limitations. Micromachines 2017, 8, 261. https://doi.org/10.3390/mi8090261.

[34] Faisal, N., Zindani, D., Kumar, K., Bhowmik, S. (2019). Laser Micromachining of Engineering Materials—A Review. In: Kumar, K., Zindani, D., Kumari, N., Davim, J. (eds) Micro and Nano Machining of Engineering Materials. Materials Forming, Machining and Tribology. Springer, Cham. https://doi.org/10.1007/978-3-319-99900-5_6.

[35] Malcolm C. Gower "Laser micromachining for manufacturing MEMS devices", Proc. SPIE 4559, MEMS Components and Applications for Industry, Automobiles, Aerospace, and Communication, (1 October 2001); https://doi.org/10.1117/12.443040.

[36] Mohammad Reza Dianati, Farshid Malek Ghaini, Mohammad Javad Torkamany. Development of melt pool profile in Nd:YAG and fiber pulsed laser spot welding of 316 stainless steel foils, Journal of Materials Research and Technology, Volume 36, Pages 4614-4624, https://doi.org/10.1016/j.jmrt.2025.04.141.

[37] L. Li, C. Achara, Chemical Assisted Laser Machining for The Minimisation of Recast and Heat Affected Zone, CIRP Annals, Volume 53, Issue 1, Pages 175-178, (2004) https://doi.org/10.1016/S0007-8506(07)60672-6.

[38] Dutta Majumdar J, Manna I. Laser material processing. International Materials Reviews. 2024;56(5-6):341-388. https://doi.org/10.1179/1743280411Y.0000000003.

[39] Hribar, L.; Gregorčič, P.; Senegačnik, M.; Jezeršek, M. The Influence of the Processing Parameters on the Laser-Ablation of Stainless Steel and Brass during the Engraving by Nanosecond Fiber Laser. Nanomaterials 2022, 12, 232. https://doi.org/10.3390/nano12020232.

[40] Le, H.; Penchev, P.; Henrottte, A.; Bruneel, D.; Nasrollahi, V.; Ramos-de-Campos, J.A.; Dimov, S. Effects of Top-hat Laser Beam Processing and Scanning Strategies in Laser Micro-Structuring. Micromachines 2020, 11, 221. https://doi.org/10.3390/mi11020221.

[41] Yilmaz, B. D.; Tepe, I. H.; et al, Effect of Yb: Fiber laser on surface roughness and wettability of titanium. Balkan Journal of Dental Medicine 26(1):52-57 (2022). http://dx.doi.org/10.5937/bjdm2201052D.

[42] El-Bassyouni, G.T., Mouneir, S.M. & El-Shamy, A.M. Advances in surface modifications of titanium and its alloys: implications for biomedical and pharmaceutical applications. Multiscale and Multidiscip. Model. Exp. and Des. 8, 265 (2025). https://doi.org/10.1007/s41939-025-00823-1.

[43] Marin, E.; Lanzutti, A. Biomedical Applications of Titanium Alloys: A Comprehensive Review. Materials 2024, 17, 114. https://doi.org/10.3390/ma17010114.

[44] Ruina Jiao, Kunlun Wang, Yanqing Xin, Hui Sun, Jianhong Gong, Lan Yu, Yong Wang, S. Enhancing the temperature coefficient of resistance of Pt thin film resistance-temperature-detector by short-time annealing, Ceramics International, Volume 49, Issue 8, Pages 12596-12603, 2023. https://doi.org/10.1016/j.ceramint.2022.12.122.

[45] K. Tsutsumi, A. Yamashita and H. Seidel, "The experimental study of high TCR Pt thin films for thermal sensors," SENSORS, 2002 IEEE, Orlando, FL, USA, 2002, pp. 1002-1005 vol.2, doi: 10.1109/ICSENS.2002.1037248.

[46] Nam VB, Lee D. Evaluation of Ni-Based Flexible Resistance Temperature Detectors Fabricated by Laser Digital Pattering. Nanomaterials (Basel). 2021 Feb 25;11(3):576. doi: 10.3390/nano11030576.

[47] M. N. Mohamed Zukri, M. S. Al Farisi, Y. Hasegawa and M. Shikida, "Single-Step Laser Fabrication of 3D Free-Standing Origami MEMS Thermal Sensor," 2024 IEEE SENSORS, Kobe, Japan, 2024, pp. 1-4, doi: 10.1109/SENSORS60989.2024.10785084.

[48] S. Moroe, P. L. Woodfield, J. Fukai, K. Shinzato, M. Kohno, M. Fujii and Y. Takata, "Thermal Conductivity Measurement of Gases by the Transient Short-Hot-Wire Method," Experimental Heat Transfer, vol. 24, no. 2, p. 168–178, 2011. https://doi.org/10.1080/08916152.2010.503310.

[49] S. Saremi, A. Alyari, D. Feili and H. Seidel, "A MEMS-based hot-film thermal anemometer with wide dynamic measurement range," Proceedings of IEEE Sensors, pp. 420-423, 2014. doi: 10.1109/ICSENS.2014.6985024.

[50] T. W. K. Jonathan, Y. Lawrence and M. Ellis, "Micromachined Thermal Flow Sensors—A Review," Micromachines, vol. 3, pp. 550-573, 2012. https://doi.org/10.3390/mi3030550.

[51] A. Manuel, d. M. Amador and V. d. S. Ferreira, "Gas mass-flow meters: Principles and applications," Flow Measurement and Instrumentation, vol. 21, no. 2, pp. 143-149, 2010. https://doi.org/10.1016/j.flowmeasinst.2010.02.001.

[52] N. Nguyen, "Micromachined flow sensors—a review," Flow Measurement and Instrumentation, vol. 8, no. 1, pp. 7-16, 1997. https://doi.org/10.1016/S0955-5986(97)00019-8.

[53] M. N. B. Mohamed Zukri, M. S. Al Farisi, Y. Hasegawa and M. Shikida, "Laser Micromachining 3-D Free Standing Structure of MEMS Thermal Flow Sensor," in IEEE Sensors Journal, vol. 25, no. 11, pp. 18876-18883, 1 June1, 2025, doi: 10.1109/JSEN.2025.3559179.

[54] M. Ohring, "11 - Electrical Properties Of Metals, Insulators, And Dielectrics," Engineering Materials Science, pp. 559-610, 1995. https://doi.org/10.1007/s10854-020-04974-4.

[55] Al Farisi, M.S., Kawata, T., Hasegawa, Y., Zukri, M.N.M., Matsushima, M., Kawabe, T. and Shikida, M., " Laser fabrication of 3-D free-standing thin sensor and facile MEMS flow sensor integration for implantable respiration monitoring," Sensors and Actuators A: Physical, p.117083, doi: https://doi.org/10.1016/j.sna.2025.117083.

[56] Surinder Singh, Alok Jejusaria, et al; Comparative study of titanium, platinum, and titanium nitride thin films for micro-electro mechanical systems (MEMS) based micro-heaters. AIP Advances 12, 095202 (2022). https://doi.org/10.1063/6.0001892.

[57] Campanelli, L. A review on the recent advances concerning the fatigue performance of titanium alloys for orthopedic applications. Journal of Materials Research 36, 151–165 (2021). https://doi.org/10.1557/s43578-020-00087-0.

[58] Kacsó AB, Peter I. A Review of Past Research and Some Future Perspectives Regarding Titanium Alloys in Biomedical Applications. J Funct Biomater. 2025 Apr 18;16(4):144. doi: 10.3390/jfb16040144.

**Mohammad Nizar Bin Mohamed Zukri** received the B.E. degree in mechanical engineering in 2015 and the M.E. degree in control systems in 2018 from the University of Science Malaysia (USM), Penang, Malaysia. In 2018, he was a collaborating researcher with the Department of Mechanical Engineering at Toyohashi University of Technology (TUT), Aichi, Japan. From 2018 to 2023, he worked in the semiconductor industry (Flextronics and Western Digital) as an engineer. His research interests include control systems and instrumentation, the development of microelectromechanical systems (MEMS), and packaging technologies. Currently, he is pursuing a Ph.D. degree in MEMS research at Hiroshima City University (HCU), Hiroshima, Japan.

**Muhammad Salman Al Farisi** received the B.E. degree in mechanical engineering, and the M.E. and Ph.D. degrees in robotics from Tohoku University, Sendai, Japan, in 2016, 2018, and 2021, respectively. In 2017, he was a Visiting Researcher with the Fraunhofer Institute for Electronic Nano Systems (ENAS), Chemnitz, Germany. From 2019 to 2021, he was a Research Fellow of the Japan Society for the Promotion of Science (JSPS) with the Department of Robotics, Tohoku University. With the Department of Biomedical Information Sciences, Hiroshima City University, he was an Assistant Professor from 2021 to 2024, and has been a Lecturer since 2024. His research interests include the development of microelectromechanical systems (MEMS) and mechatronics, in particular for biomedical applications.

**Yoshihiro Hasegawa** received the M.S. and Ph.D. degrees in engineering from Nagoya University, Nagoya, Japan, in 2005 and 2008, respectively. From 2008 to 2015, he worked at Canon Inc., Tokyo, Japan. He is currently working as an Associate Professor with the Department of Biomedical Information Sciences, Hiroshima City University, Hiroshima, Japan. His research interests include microsensors for robotics and medical applications. Dr. Hasegawa is a member of the Japan Society of Mechanical Engineers.

**Mitsuhiro Shikida** received the B.S. and M.S. degrees in electrical engineering from Seikei University, Tokyo, Japan, in 1988 and 1990, respectively, and the Ph.D. degree from Nagoya University, Nagoya, Japan, in 1998. He was with Hitachi Ltd., Tokyo, from 1990 to 1995, and Nagoya University from 1995 to 2014. In 2014, he joined the Department of Biomedical Information Sciences, Hiroshima City University, Hiroshima, Japan, as a professor. He has published over 140 peer-reviewed articles, over 200 presentations at international conferences, and 22 patents. His current research interests include the integration of microsensors for intelligent systems and their biomedical applications. Dr. Shikida was an Editorial Board Member of Micro and Nano Technology (IET) and Sensors (MDPI). He was a Topics Editor of Sensors and Actuators A: Physical (Elsevier). He has served on the Technical Program Committee at the International Conference, IEEE-MEMS, Transducers, and IEEE-Sensors.